\documentclass[journal=apchd5,manuscript=article ]{achemso}
\usepackage[utf8]{inputenc}
\usepackage[T1]{fontenc}
\usepackage{mathptmx}
\usepackage{etoolbox}
\usepackage{xcolor}
\usepackage{booktabs} 
\usepackage{tabularx}
\usepackage{graphicx}
\usepackage{dcolumn}
\usepackage{bm}
\usepackage{setspace}
\usepackage{soul}  
\usepackage{lineno}

\usepackage{siunitx} 
\usepackage[version=3]{mhchem} 

\author{Jacopo Pelini}

    \affiliation{CNR-INO -- Istituto Nazionale di Ottica, Via Carrara, 1 -- 50019 Sesto Fiorentino FI, Italy.}
    \alsoaffiliation{LENS -- European Laboratory for Non-Linear Spectroscopy, Via Carrara, 1 -- 50019 Sesto Fiorentino FI, Italy.}

\author{Stefano Dello Russo}

    \affiliation{ASI, Italian Space Agency, Località Terlecchia, Matera, 75100, Italy.}

\author{Chenghong Zhang}
    \affiliation{CNR-INO -- Istituto Nazionale di Ottica, Via Carrara, 1 -- 50019 Sesto Fiorentino FI, Italy.}
    \alsoaffiliation{LENS -- European Laboratory for Non-Linear Spectroscopy, Via Carrara, 1 -- 50019 Sesto Fiorentino FI, Italy.}

\author{Zhen Wang}

    \affiliation{Harbin Institute of Technology, Shenzhen, China.}

\author{Iacopo Galli}

    \affiliation{CNR-INO -- Istituto Nazionale di Ottica, Via Carrara, 1 -- 50019 Sesto Fiorentino FI, Italy.}
    \alsoaffiliation{LENS -- European Laboratory for Non-Linear Spectroscopy, Via Carrara, 1 -- 50019 Sesto Fiorentino FI, Italy.}

\author{Pablo Cancio Pastor}

    \affiliation{CNR-INO -- Istituto Nazionale di Ottica, Via Carrara, 1 -- 50019 Sesto Fiorentino FI, Italy.}
    \alsoaffiliation{LENS -- European Laboratory for Non-Linear Spectroscopy, Via Carrara, 1 -- 50019 Sesto Fiorentino FI, Italy.}

\author{Maria Concetta Canino}

    \affiliation{CNR-ISMN, The Institute for the Study of Nanostructured Materials of the National Research Council of Italy, Via P. Gobetti 101, Bologna, 40129, Italy.}

\author{Alberto Roncaglia}

    \affiliation{CNR-ISMN, The Institute for the Study of Nanostructured Materials of the National Research Council of Italy, Via P. Gobetti 101, Bologna, 40129, Italy.}

\author{Naota Akikusa}

    \affiliation{Hamamatsu Photonics K.K, Shizuoka, Japan.}

\author{Wei Ren}

    \affiliation{Department of Mechanical and Automation Engineering, The Chinese University of Hong Kong, New Territories, Hong Kong SAR, China.}

\author{Mario Siciliani de Cumis}
    \affiliation{ASI, Italian Space Agency, Località Terlecchia, Matera, 75100, Italy.}
    \alsoaffiliation{CNR-INO -- Istituto Nazionale di Ottica, Via Carrara, 1 -- 50019 Sesto Fiorentino FI, Italy.}
        \email{mario.sicilianidecumis@asi.it}

\author{Paolo De Natale}

    \affiliation{CNR-INO -- Istituto Nazionale di Ottica, Via Carrara, 1 -- 50019 Sesto Fiorentino FI, Italy.}
    \alsoaffiliation{LENS -- European Laboratory for Non-Linear Spectroscopy, Via Carrara, 1 -- 50019 Sesto Fiorentino FI, Italy.}

\author{Simone Borri}
    \affiliation{CNR-INO -- Istituto Nazionale di Ottica, Via Carrara, 1 -- 50019 Sesto Fiorentino FI, Italy.}
    \alsoaffiliation{LENS -- European Laboratory for Non-Linear Spectroscopy, Via Carrara, 1 -- 50019 Sesto Fiorentino FI, Italy.}

\title
  {Sub-doppler trace-gas photoacoustic spectroscopy}


\begin{document}

\singlespacing

\begin{abstract}
Molecules are emerging as new benchmark for metrology and fundamental physics research, driving the demand for spectroscopic techniques combining high sensitivity and resolution. Photoacoustic spectroscopy has proven to combine high sensitivity with appealing features like compactness, wavelength-independent and background-free detection. To date, photoacoustic sensing has mostly been focused on high-pressure applied trace-gas analysis, while accessing the low-pressure regime has been considered not compatible with efficient acoustic wave propagation. However, sensing gas samples at low pressure is the key to get access to high-resolution spectroscopy. Here, we demonstrate that sub-Doppler saturation spectroscopy can be performed on low-pressure trace gases in a cavity-enhanced photoacoustic sensor with mW-level mid-infrared radiation. Moreover, we show that the same setup can be operated at higher pressure, enabling trace-gas detection with \SI{5} parts-per-billion sensitivity with a laser power as low as \SI{35}{\uW}. This allows to extend the unique advantages of the photoacoustic technique to metrology and fundamental physics and provides the mid-infrared with a cost-effective, flexible tool combining high sensitivity and resolution.
\end{abstract}

\section{Introduction}

Photoacoustic (PA) sensing is undergoing an overwhelming development for trace-gas molecular detection. Indeed, based on the detection of a sound wave generated following the radiation interaction with the target molecules~\cite{bell1880photophone,rosencwaig1980photoacoustic}, this technique exploits an acousto-mechanical transduction instead of a photo-detection. This, in turn, makes this technique very flexible, because it separates the molecular excitation, for which laser frequencies resonant with the strongest fundamental absorption lines can be chosen, from detection, which is totally wavelength-independent~\cite{sun2024multicomponent,qiao2024ultra}. Moreover, since the signal is detected only as a consequence of the radiation-sample interaction and the subsequent pressure wave generation, this technique is intrinsically background-free. This is the key to achieving dynamical ranges that are completely out of reach for different direct-absorption-based techniques. One of the most impressive results in this sense has been recently demonstrated, showing an unprecedented 8-decade dynamic range achieved with a quartz-enhanced photoacoustic sensor~\cite{wang2022doubly}. 

For acoustic detection, several different transducers have been developed, spanning from conventional microphones~\cite{haisch2011photoacoustic,meng2025multicomponent} and quartz tuning forks (QTFs) ~\cite{kosterev2002quartz,wang2025ultrahigh}, which are appealing for their electrical readout, to cantilevers, which can rely on a very sensitive optical interferometric readout~\cite{wilcken2003optimization}. In our recent work, we have shown that customized cantilevers, coupled with dual-tube acoustic resonators, can be operated at high-order resonant modes, thus improving the detection sensitivity below 1-part-per-billion (ppb) minimum detection limit (MDL) in standard single-passage operation~\cite{russo2024dual}. In addition, we demonstrated that cantilever performances can be tailored, in terms of physical parameters like shape, size and mass, depending on the specific experimental requirements~\cite{pelini2024new}. Furthermore, since the PA signal is directly proportional to the optical power interacting with the molecular sample~\cite{patimisco2014quartz}, the final detection sensitivity can be further boosted by coupling the PA sensor with a high-finesse optical resonator~\cite{borri2014intracavity}. Cavity-enhanced PA sensors using cantilevers, microphones, or QTFs have been reported, achieving MDLs spanning from 10 parts-per-trillion (ppt)~\cite{pelini2025cavity} down to the sub-ppt level~\cite{wang2022doubly,nie2024sub,tomberg2019cavity}. 
 
The cavity enhancement configuration is also the key to access the low gas pressure regime. Indeed, optical resonators allow for enhancing the signal resonant with the target molecular line, generating stronger acoustic waves. In turn, this can compensate for the increasing losses experienced by acoustic signals when progressively decreasing the gas pressure. Therefore, choosing the best trade-off between pressure and detected signal enables access to the 10$\mathrm{^2}$~Pa gas pressure regime that allows observation of molecular saturated-absorptions. The possibility to access this regime was suggested in an old, pioneering work, where Doppler-free molecular lines were recorded with a bulky PA setup using a high-power fixed-wavelength \ce{CO2} laser shone on pure \ce{CO2} and \ce{CH3OH} gas samples~\cite{di1979sub}. Despite the significant technological improvement over the last decades, this work has never had a legacy, yet. 

Here, we demonstrate that, using a tailored cantilever-based cavity-enhanced PA setup, both high sensitivity and sub-Doppler resolution can be achieved on a trace-gas sample using a low-power mid-infrared (IR) Quantum Cascade Laser (QCL) by selecting the proper experimental parameters. We show Doppler-free signals with a linewidth as narrow as \SI{6}{MHz}, at total gas pressures down to \SI{60}{Pa}. This result effectively fills the last important gap between PA and the best-performing laser spectroscopy techniques used for the most demanding metrological and fundamental applications~\cite{maddaloni2016laser}, enhancing the potential impact of high-resolution spectroscopy in the molecular fingerprint region. Moreover, with this setup we demonstrate that, only changing the gas pressure, a sensitivity as high as \SI{5}{ppb}, for a \ce{N2O} in \ce{N2} sample, using an optical power as low as \SI{35}{\uV}, corresponding to a NNEA of 2.93$\times$10$\mathrm{^{-11}}$ cm$\mathrm{^{-1}}$WHz$\mathrm{^{-\frac{1}{2}}}$ can be achieved at the same time. This allows our sensor to be operated with low-power-consuming, millimeter-sized lasers while still maintaining the sensitivity required by most applications targeting environmental monitoring, human safety, and security.

\section{Experimental setup}

\begin{figure}[b!]
\centering\includegraphics[width=1.0\columnwidth]{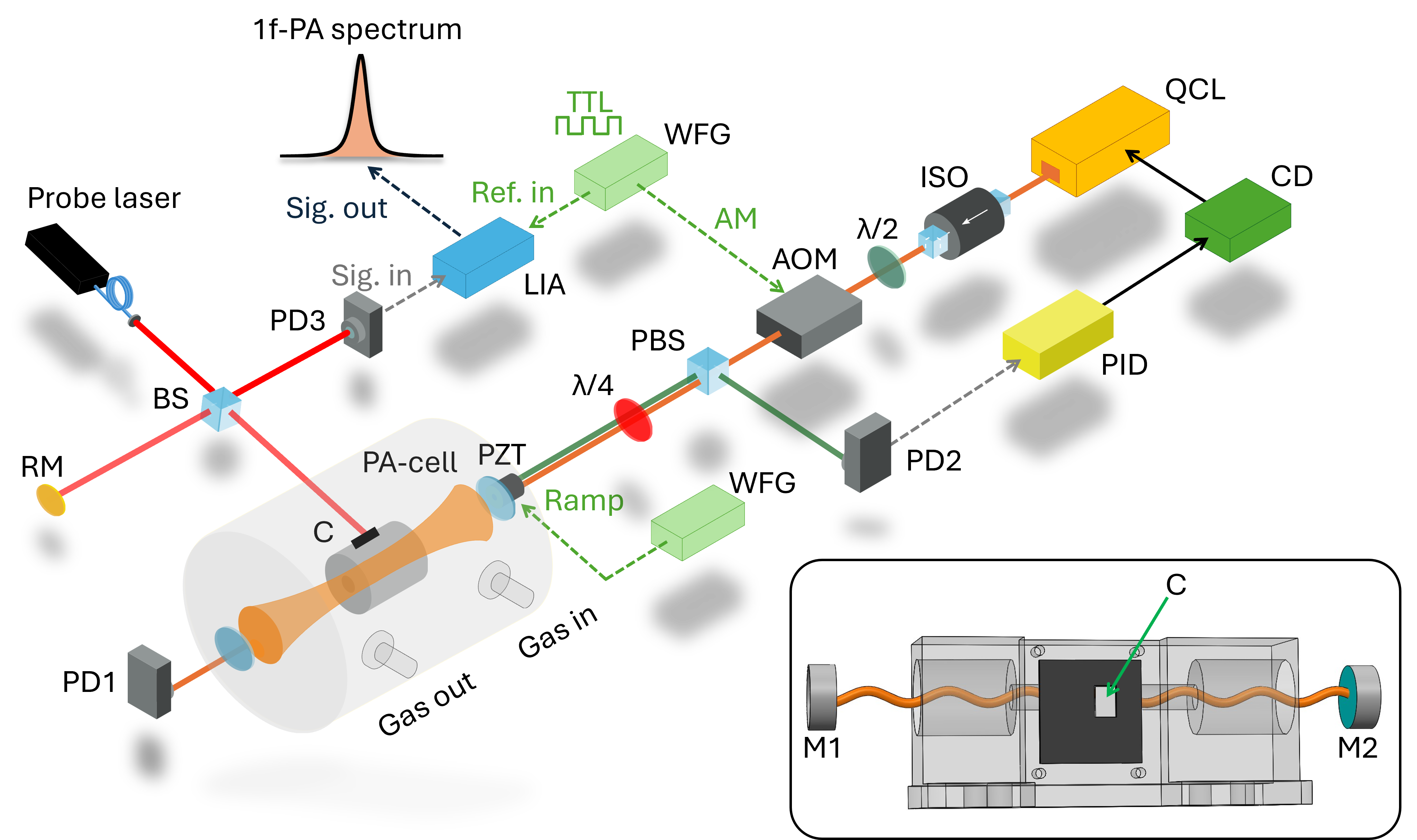}
\caption{Schematic representation of the experimental setup. CD: current driver. QCL: Quantum Cascade Laser. $\mathrm{\lambda}$/2: half wave-plate. AOM: acousto-optic modulator. PBS: polarizing beam splitter. $\mathrm{\lambda}$/4: quarter wave-plate. PZT: piezo-electric actuator. C: silicon-based cantilever. PD1: cavity transmission photo-detector. PD2: cavity reflection photo-detector. PID: proportional-integral-derivative servo controller. BS: beam splitter. RM: reference mirror. PD3: Michelson interferometer photo-detector. LIA: lock-in amplifier. WFG: waveform generator. 1f-PA spectrum: first-harmonic cantilever-enhanced photo-acoustic spectrum. \textbf{Inset}: 3D-sketch of the implemented doubly resonant configuration. The custom-made acoustic resonator allows the acoustic wave intensity amplification, while the optical cavity (consisting of the two plano-convex mirrors M1 and M2) allows the optical power enhancement. 
\label{exp_setup}}
\end{figure}

Figure~\ref{exp_setup} reports the scheme of our cantilever-based photoacoustic sensor. A mid-infrared, continuous-wave (CW) Quantum Cascade Laser (QCL) acts as the excitation source, addressing a strong  Nitrous Oxide ($\mathrm{N_2O}$) rovibrational transition with a central wavelength of \SI{4.567}{\um} and a line-strength equal to 2.14$\times$10$\mathrm{^{-19}}$ cm/molecule. All the laser parameters, i.e., the bias current and the working temperature (the latter stabilized at 20 $^\circ\mathrm{C}$), are managed by a low-noise modular controller provided by ppqSense, enabling the QCL emission with a free-running sub-MHz linewidth. An optical isolator (ISO, COHERENT FM2 MID-IR \SI{4.55}{\um}) prevents optical feedback, while an acousto-optic modulator (AOM, ISOMET, M1208-G80-4) is exploited for the optical beam modulation. Combining an 80-MHz RF and a digital modulation (TTL) at a given frequency $f_0$, the AOM periodically deflects the interacting laser beam, resulting in an intensity modulation at $f_0$ and allowing the photoacoustic effect within the PA-cell, filled with the sample gas at the desired pressure. 

The acoustic waves generated by the trace-gas non-radiative relaxation are collected via a \SI{4}{mm} long, \SI{6}{mm} high, and \SI{10}{\um} thick rectangular silicon-based cantilever (C in Figure \ref{exp_setup}), acting as a sensitive transducer \cite{pelini2024new}. The cantilever is coupled with a custom-made acoustic resonator, whose geometry is optimized for an efficient acoustic wave intensity amplification centered at the transducer's resonance frequency. A more detailed description of the cantilever and acoustic resonator's architecture can be found in the Supporting Information Section. In addition, a \SI{85}{\mm}-long Fabry-Pérot optical cavity provides an amplification of the optical power available for the photoacoustic effect generation. It consists of two plano-convex mirrors (M1 and M2) with radii of curvature of \SI{125}{\mm} and reflectivities of 0.9984, leading to a cavity finesse equal to $\simeq$\SI{1960} and an optical power enhancement of $\simeq$\SI{625}. The combination of both acoustic and optical amplification strategies allows the achievement of a doubly-resonant configuration. As reported in the inset of Figure~\ref{exp_setup}, the optical cavity is coaxial with the acoustic resonator, and the cantilever is mounted on a small aperture at the resonator mid-point. In this way, the system is designed to enable the best possible superposition of the cavity beam waist, the acoustic wave amplitude, and the cantilever's displacement. This particular configuration allows the minimization of the interacting optical area and, simultaneously, the maximization of the acoustic wavefronts impacting the cantilever's free end. 

In our setup,  the intra-cavity optical power is stabilized by locking the laser to one of the cavity modes through a Pound-Drever-Hall (PDH) feedback loop~\cite{black2001introduction}. The cavity reflection (depicted in green) is picked up by a quarter-wave plate ($\mathrm{\lambda/4}$) and a polarizing beam splitter (PBS). The signal is detected by the photo-detector PD2 (Vigo Systems, PVI-4TE-5-1x1-TO8). The beat note between the laser carrier and side bands (the latter generated via a 4-MHz local oscillator) is demodulated by a mixer to retrieve the error signal. After being processed by a PID servo controller, the correction signal is sent to the laser driver to maintain the QCL locked with the optical cavity. The photo-detector PD1 (Vigo Systems, PVI-4TE-5-1x1-TO8), placed after the output cavity mirror, serves only for alignment purposes.

The photoacoustic signal is retrieved by optically converting the cantilever's oscillation via a Michelson interferometric readout~\cite{pelini2025cavity,patentPAS}. A HeNe probe laser is split into two identical paths via a 50/50 beam splitter (BS): one enters the PA-cell and impinges on the cantilever's free end, while the other is sent to a reference mirror. The two back-reflected paths are recombined at the same beam splitter and then sent to a photo-detector (PD3). The interferometric output is demodulated at the cantilever's resonance frequency via a Lock-in amplifier (LIA, MFLI, Zurich Instruments 500kHz/5MHz), thus retrieving the in-phase first harmonic R-component of the signal. The spectral scan over the absorption line is achieved by applying a slow ramp ($f$ = 2 mHz) to a PZT attached to the input cavity mirror. All the analysis is performed in static conditions on highly diluted \ce{N2O}:\ce{N2} samples.

\section{Results and Discussion}\label{sec3}

\subsection{Cantilever's mechanical response}\label{subsect3.1}

\begin{figure}[h!]
\centering\includegraphics[width=1\textwidth]{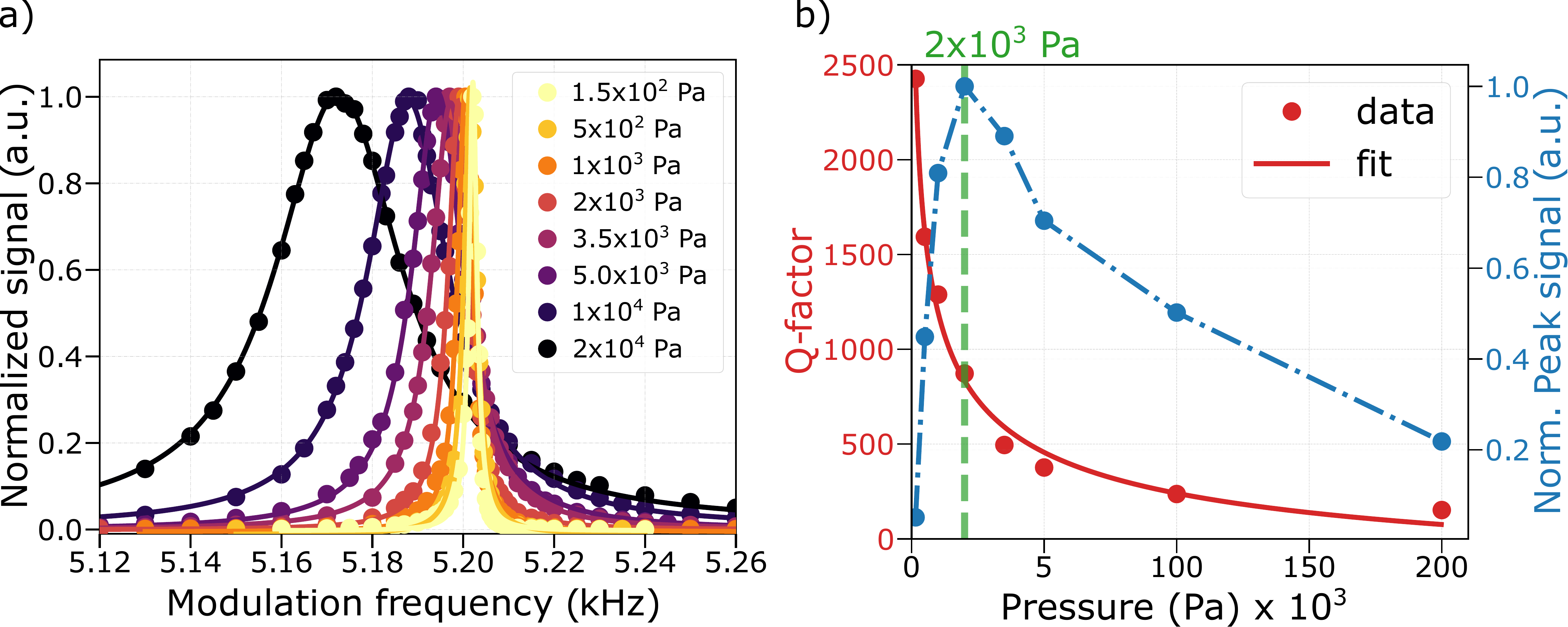}
\caption{\textbf{(a)}: Normalized resonance curve spectra of the cantilever at different working pressures ranging from 1.5$\mathrm{\times 10^2}$~PA to 2$\mathrm{\times 10^4}$~Pa. The solid curves represent the Lorentzian-type fit on each experimental data set. \textbf{(b)}: Quality factor (red data points) and normalized peak signal value (blue data points) versus the working pressure.
\label{Qfact_vs_P}}
\end{figure}

As a first, fundamental step, the mechanical properties of the acoustic transducer were characterized. From a theoretical point of view, a Finite Element Method (FEM) was used to perform the cantilever's eigenfrequencies analysis, simulating the resonance oscillation frequencies and the associated resonant mode's displacement (i.e., the eigenmode). The results of the simulation process, limited to the first eight modes, are reported in the Supplementary Materials. In this work, the fifth resonance mode was employed. Being characterized by an estimated frequency of \SI{5577.25}{Hz}, this choice represents a good compromise between sufficiently high immunity from low-frequency environmental acoustic and mechanical noise and efficient sound wave detection and conversion~\cite{pelini2024new}. The cantilever's mechanical properties were studied by exploiting a photoacoustic measurement on gas samples at different pressures, ranging from 1.5$\mathrm{\times 10^2}$ Pa to 2$\mathrm{\times 10^3}$ Pa. For each working pressure, the R$\mathrm{^2}$-component of the photoacoustic signal was stepwise acquired, varying the AOM modulation frequency. The latter signal, defined as the resonance curve spectrum, was fitted with a Lorentzian-type peak function. The central frequency (i.e., the experimental cantilever's resonance frequency $f_0$)  and the spectral Full-Width-at-Half-Maximum ($\Delta f$) were extracted as fit parameters, and the quality factor was computed as $Q=f_0/\Delta f$.

According to Figure~\ref{Qfact_vs_P}\textbf{(a)}, as the pressure increases, the cantilever's resonance curve spectrum experiences an FWHM-broadening and a decrease in the associated resonance frequency. This effect arises from the increase in the viscous damping effects of the surrounding molecules, leading to a greater energy dissipation in the interacting oscillating body. The combination of these two phenomena leads to progressive degradation of the mechanical quality factor with increasing pressure (red data points in Figure \ref{Qfact_vs_P}\textbf{(b)}),  following the typical trend proportional to $1/\sqrt{P}$~\cite{blom1992dependence,patimisco2016analysis} depicted with the red solid line. As it is well-known, the overall PA response results from the interplay of two different mechanisms. On one side, the optimal cantilever mechanical response (i.e., its quality factor) is found at low sample gas pressure. On the other side, however, an efficient generation and propagation of acoustic waves in the gas sample requires a sufficiently high number of molecules and so higher pressures. The combination of these two contributions determines the overall photoacoustic response. The blue data points in Figure \ref{Qfact_vs_P}\textbf{(b)} depict the PA peak signal amplitude as a function of the pressure, acquired in the single-passage configuration with a concentration of \SI{10}{ppm} \ce{N2O} diluted in \ce{N2}. The optimal working pressure of 2$\mathrm{\times 10^3}$ Pa represents a trade-off between the two competing contributions mentioned above. The sharp drop in the PA signal at lower pressures clearly explains why working at low-pressure regimes is challenging with PA systems and, therefore, usually unpracticed.

\subsection{Accessing the sub-Doppler regime}

\begin{figure}[h!]
\centering\includegraphics[width=1\textwidth]{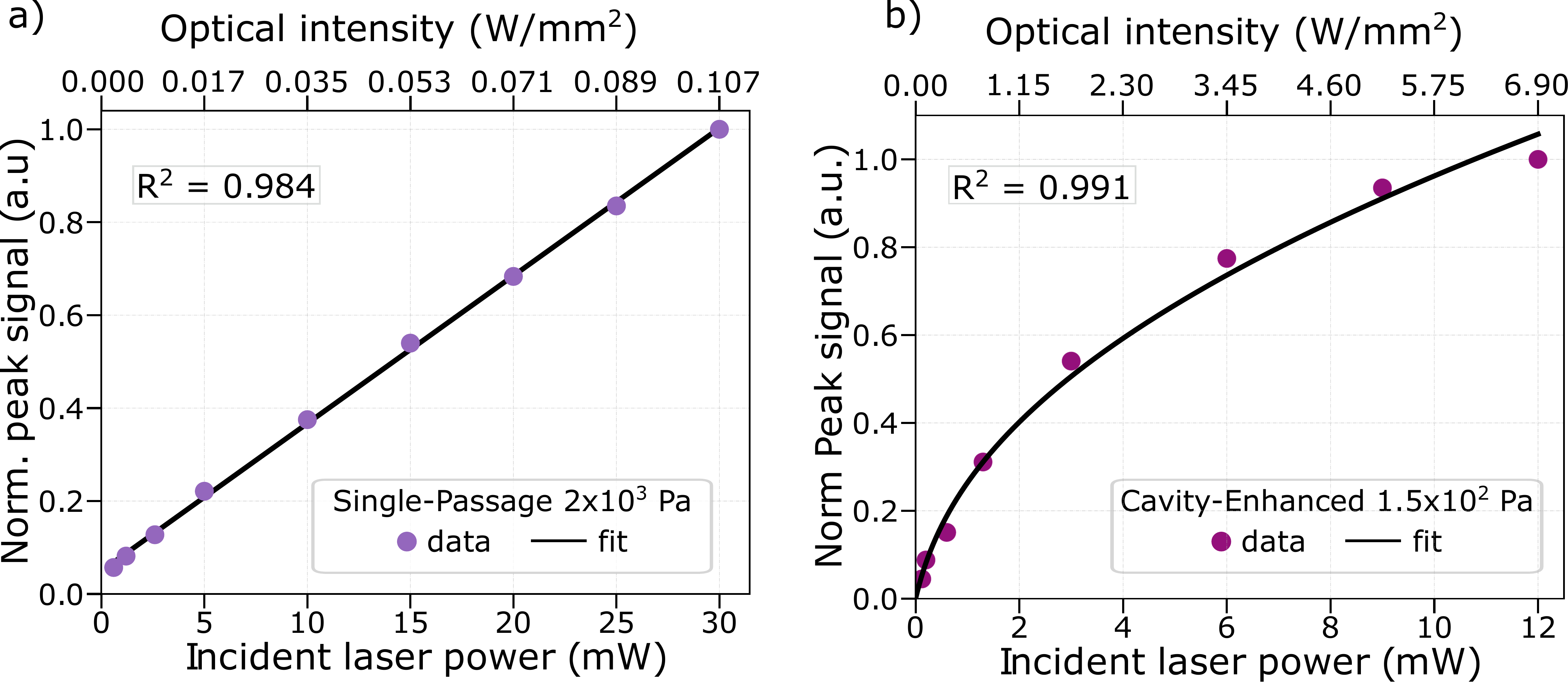}
\caption{\textbf{(a)}: Normalized PA peak signal as a function of the incident laser power and of the optical intensity in the single-passage configuration at the best working pressure of 2$\mathrm{\times 10^3}$ Pa. \textbf{(b)}: Normalized PA peak signal versus the laser incident power and the optical intensity in the cavity-enhanced configuration at a pressure of 1.5$\mathrm{\times 10^2}$ Pa.
\label{SvsP}}
\end{figure}

When a PA system is combined with a high-finesse optical cavity, the intra-cavity power enhancement strongly boosts the PA signal, thus influencing the overall response of the system. It not only allows for increasing the detection sensitivity, as already widely demonstrated~\cite{borri2014intracavity,patimisco2015high,nie2023mid,zheng2024high,qiao2024ultra}, but also can allow the investigation of low-pressure gas samples (1$\mathrm{\times 10^2}$~Pa or below), thus unveiling nonlinear-absorption effects.

Figures \ref{SvsP}\textbf{(a)} and \ref{SvsP}\textbf{(b)} show the normalized PA peak signal trends for the single-passage and cavity-enhanced configurations, respectively, as a function of the laser incident power. In the single-passage configuration and at 2$\mathrm{\times 10^3}$~Pa sample pressure, the signal linearly increases with the power, confirming the well-known behavior~\cite{patimisco2014quartz,haisch2011photoacoustic}. When the optical cavity comes into play and the working pressure is reduced down to 1.5$\mathrm{\times 10^2}$~Pa, the acquired PA signal experiences a non-linearity as a consequence of \ce{N2O} absorption line saturation effects due to the high intra-cavity power buildup. 

\newpage

For a quantitative analysis of the saturation effect, we must start considering the line-broadening contribution in our experimental conditions. The addressed \ce{N2O} rovibrational transition is affected by three main line-broadening mechanisms. The first one is the Doppler-limited broadening, which depends on the incident radiation frequency shift due to the Doppler effect. It leads to a FWHM broadening equal to $\Delta_D = 2k_0\sqrt{\frac{2k_BTln2}{mc^2}}$, where $m$ is the molecule's mass ($\mathrm{N_2O}$, \SI{44.01}{g/mol}), $c$ the light speed, $k_B$ the Boltzmann constant, and $T$ the temperature. The second is related to collisional effects. It is described by the pressure-dependent coefficient $\Delta_p = C_pp$, where $C_p$ [MHz Pa$^\mathrm{^{-1}}$] is the broadening coefficient (obtainable from the HITRAN database of the given transition~\cite{rothman2013hitran2012}), and $p$ is the working pressure. The third contribution comes from the transit time effect~\cite{gotti2018comb}. It is equal to  $\Delta_t = \overline{v}/4\pi {R_c}$, where $\overline{v}=\sqrt{\frac{8k_BT}{\pi m}}$ is the averaged molecule's speed at a temperature $T$, and $R_c$=\SI{0.3}{\mm} is the beam waist at the center of the optical cavity. At 1.5$\mathrm{\times 10^2}$~Pa, considering our experimental conditions, these three effects contribute with FWHMs equal to $\Delta_D\simeq$\SI{122.7}{MHz}, $\Delta_p\simeq$\SI{3.1}{MHz}, and $\Delta_t\simeq$\SI{1.8}{MHz}, respectively. It follows that, at this range of pressure, the inhomogeneous broadening due to the Doppler effect is predominant. In this condition, the absorption signal $S(I)$ is described by: 

\begin{equation}\label{eq_1}
    S(I) = \frac{S_0 I}{\sqrt{\left(1+\frac{I}{I_s}\right)}}, 
\end{equation}

where $I$ is the optical intensity, and $I_s$ is the expected saturation intensity for the addressed transition, given by:

\begin{equation}\label{eq_2}
    I_S =\frac{\hbar\omega_{LAS}}{2\pi c^2 A_E}\times(2\pi\Delta_p+2\pi\Delta_t)^2.
\end{equation}

Here $\omega_{LAS}=2\pi f_{LAS}$ is the laser angular frequency, $\hbar$ the Planck constant, and $A_E$ the Einstein coefficient of the transition. Given the values of these physical parameters in this experimental working condition, at 1.5$\mathrm{\times 10^2}$~Pa the \ce{N2O} rovibrational transition shows a saturation intensity $I_s\simeq$\SI{0.12}{W/mm}$\mathrm{^2}$. According to this analysis, the experimental data reported in Figure~\ref{SvsP}\textbf{(b)} are fitted with Equation~\ref{eq_1}. The upper x-axis, reporting the intra-cavity optical intensity, is converted from the laser power incident on the cavity according to:

\begin{equation}\label{eq_3}
    I_{ic} = G_{ic}\times \frac{P_{in}}{\pi\times R_c^2},
\end{equation}

where is the QCL-cavity coupling efficiency, $R_c$ is the QCL beam waist radius at the center of the optical cavity, and $G_{ic}$ is the intra-cavity build-up factor in the experimental working conditions. As can be seen, the fit well reproduces the experimental data. 

Accessing the saturated absorption regime with a sufficiently high signal at low pressure is the fundamental condition to unfold challenging and unconventional applications for PA-based systems, such as high-resolution spectroscopy measurements on trace-molecule samples. Proof of this assumption is the experimental observation of a Lamb dip over the absorption spectral line.
\begin{figure}[h!]
\centering\includegraphics[width=1\columnwidth]{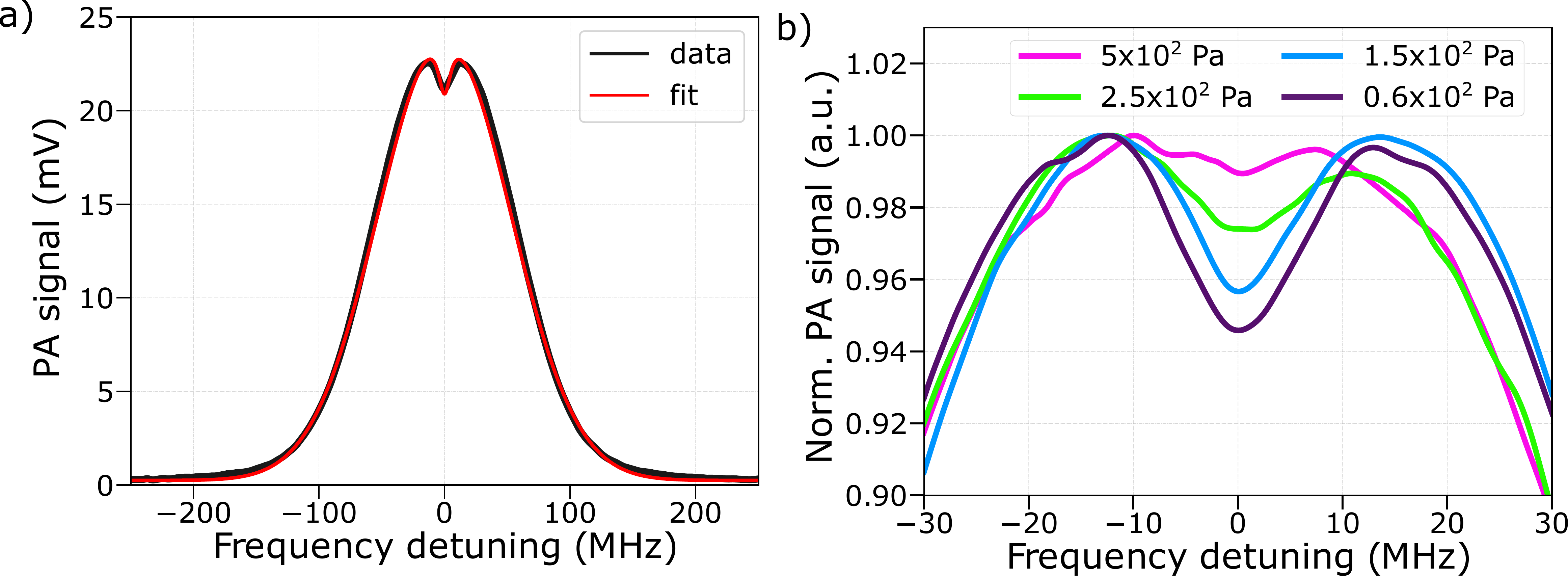}
\caption{\textbf{(a)}: 1f-PA spectrum acquired at 1.5$\mathrm{\times 10^2}$~Pa and with an incident laser power of \SI{0.7}{mW}. In this condition, the measured Lamb dip contrast is the highest, being equal to 13.03\%. \textbf{(b)}: Normalized 1f-PA spectra at different working pressures from 5$\mathrm{\times 10^2}$~Pa to 0.6$\mathrm{\times 10^2}$~Pa, at a fixed incident laser power of \SI{6}{mW}. 
\label{Lamb_dips}}
\end{figure}

The black trace in Figure \ref{Lamb_dips}\textbf{(a)} reports the 1f-PA spectrum obtained from a gas sample of \SI{200}{ppm} \ce{N2O} diluted in \ce{N2}, at a pressure of 1.5~$\mathrm{\times 10^2}$~Pa and with an incident laser power of \SI{0.7}{mW}. The Lamb dip is clearly visible at the center of the spectral line. The overall spectrum was fitted with a Lorentzian-type shape over the Doppler-broadened Gaussian profile. The result of the fit, reported by the solid red line in Figure~\ref{Lamb_dips}\textbf{(a)}, allows us to extract a Lamb dip contrast of 13.03\%, very close to the maximum achievable for intra-cavity pump-probe schemes (i.e., 13.3\%~\cite{letokhov1977nonlinear}), and a Lamb dip FWHM equal to (6.2~$\pm$~0.2)~MHz. For comparison, the expected Lamb dip FWHM is $\Delta = \left(\sqrt{\Delta_p^2+\Delta_t^2}\right)\sqrt{1+S} = $ \SI{5.02}{MHz}, where the pressure and transit time broadenings ($\Delta_p$ and $\Delta_t$) and the power broadening contribution ($\sqrt{1+S}$) are considered. Here, $S=$~0.95 is the saturation parameter in the measurement experimental condition. Figure~\ref{Lamb_dips}\textbf{(b)} reports 1f-PA spectra acquired at a fixed \SI{6}{mW} incident power and different sample pressures. As expected, the Lamb dip contrast rapidly decreases with increasing pressure, becoming barely visible as early as 5$\mathrm{\times 10^2}$~Pa. 

\subsection{High sensitivity performance of the PA setup}

\begin{figure}[h!]
\centering\includegraphics[width=1\textwidth ]{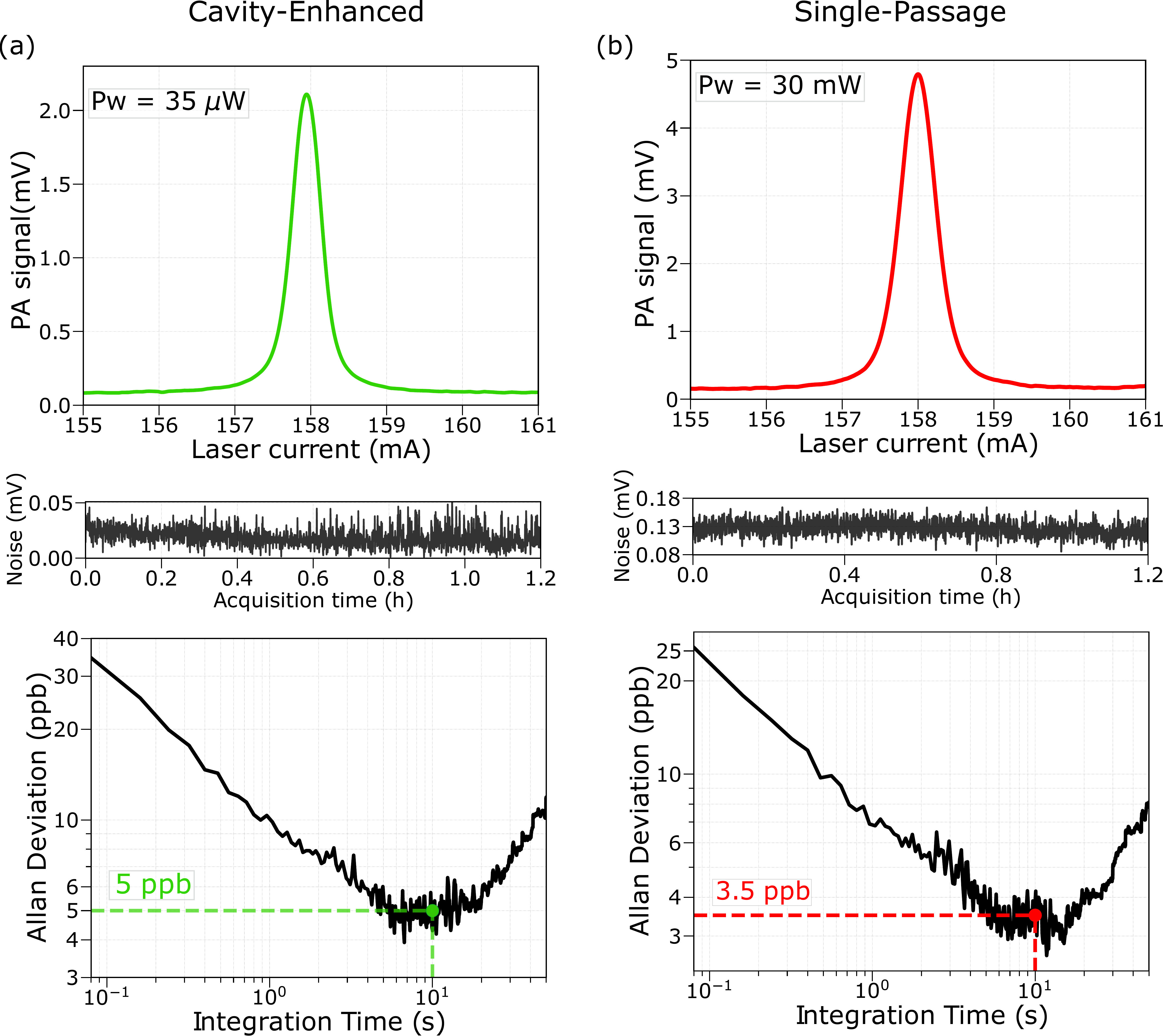}
\caption{\textbf{(a)}: 1f-PA spectrum (top panel), long-term noise trace (central panel), and Allan-Werle deviation analysis (bottom panel) in the cavity-enhanced configuration at 2 $\mathrm{\times 10^3}$ Pa of pressure and \SI{35}{\uW} of incident laser power. \textbf{(b)}: 1f-PA spectrum (top panel), long-term noise trace (central panel), and Allan-Werle deviation analysis (bottom panel) in the single-passage configuration at 2 $\mathrm{\times 10^3}$ Pa of pressure and \SI{30}{\mW} of incident laser power.
\label{comparison_Signals_20mbar}}
\end{figure}

To prove the dual-use $-$resolution and sensitivity$-$ of our sensor, we performed PA trace-gas measurements at the best pressure condition. (i.e., 2~$\mathrm{\times 10^3}$~Pa). We operated our cavity-enhanced setup with strongly reduced laser power (\SI{35}{\uW}), and we compared the achieved performance by repeating the measurements with the same sensor in a standard single-passage configuration at full laser power (\SI{30}{\mW}). Figure~\ref{comparison_Signals_20mbar}\textbf{(a)} summarizes the results in the cavity-enhanced configuration, showing an example of a 1f-PA spectrum (upper panel), a long-term noise acquisition (central panel), and the corresponding Allan-Werle deviation analysis result (bottom panel). The optical power incident onto the cavity was reduced to \SI{35}{\uW} by using an optical attenuator. In this condition, the intra-cavity power intensity is far from the saturation regime. The 1f-PA spectrum was recorded with a \SI{10}{ppm} \ce{N2O}:\ce{N2} sample. The long-term noise was measured by filling the chamber with pure \ce{N2} (at 2 $\mathrm{\times 10^3}$ Pa) to prevent any residual interaction between the target molecules and the excitation laser beam. According to the measurements, and thanks to the intra-cavity power buildup, a ppb-level detection limit was reached ($\simeq$\SI{5}{ppb}), with 10~seconds averaging time. As shown in Figure~\ref{comparison_Signals_20mbar}\textbf{(b)}, in order to retrieve a similar sensitivity level, the standard single-passage PA sensor must be operated at full laser power (\SI{30}{mW}). Both measurements were performed with the same trace-gas sample conditions. 

To quantitatively evaluate the improvement offered by the optical cavity, the detection limits were normalized by the emitted powers, thus obtaining the Noise Equivalent Absorption coefficients (NNEAs). Starting from a NNEA value equal to 1.86$\times$10$\mathrm{^{-8}}$~cm$\mathrm{^{-1}}$WHz$\mathrm{^{-\frac{1}{2}}}$ in the single-passage configuration, the optical cavity allows an NNEA reduction down to 2.93$\times$10$\mathrm{^{-11}}$, resulting in an enhancement in sensitivity of $\simeq$630, in agreement with the expected intra-cavity build-up factor. 

Finally, in Table \ref{Tab_comparison}, our result is compared with state-of-the-art photoacoustic measurements on \ce{N2O}, within the same \SI{4.5}{}-\SI{4.6}{\um} spectral range. This comparison shows that the best NNEA of 2.93$\times$10$\mathrm{^{-11}}$~cm$\mathrm{^{-1}}$WHz$\mathrm{^{-\frac{1}{2}}}$is achieved with our setup, although the best MDL is observed, as expected, at several orders of magnitude higher laser power and pressure values.  

\begin{table*}[h]
\caption{State-of-the-art photoacoustic-based sensors for \ce{N2O} detection. In the last row, the result achieved in this work in the cavity-enhanced configuration is reported. \label{Tab_comparison}}

\newcolumntype{C}{>{\centering\arraybackslash}X}
\begin{tabularx}{\textwidth}{CCCCC}
\toprule[2pt]
Wavelength ($\mathrm{\mu}$m) & Pressure (Pa) & Power (mW) & MDL and integration time & NNEA (cm$\mathrm{^{-1}}$WHz$\mathrm{^{-\frac{1}{2}}}$)\\
\midrule[2pt]

 4.56\cite{yang2024highly} & 1$\mathrm{\times10^5}$& 15 & 0.25ppb at 210s& 1.50$\times$10$\mathrm{^{-8}}$ \\
 
 4.53\cite{cao2021humidity} & 1$\mathrm{\times10^5}$ & 17.5& 1ppb at 600s &  5.69$\times$10$\mathrm{^{-9}}$\\

 4.52\cite{zifarelli2022multi} & 6$\mathrm{\times10^4}$& 25 & 7ppb at 0.1s & 5.40$\times$10$\mathrm{^{-9}}$ \\

 4.61\cite{ma2013qepas} & 1.33$\mathrm{\times10^4}$ & 987 & 23ppb at 1s& 2.90$\times$10$\mathrm{^{-9}}$\\

  4.56\cite{russo2024dual} & 2$\mathrm{\times10^4}$ & 16 & 0.34ppb at 10s& 1.3$\times$10$\mathrm{^{-9}}$\\

 4.59\cite{hayden2022mid} & 4.9$\mathrm{\times10^4}$ & 19.5 & 0.75ppb at 10s&1.70$\times$10$\mathrm{^{-10}}$ \\

4.56\cite{pelini2025cavity} & 3$\mathrm{\times10^4}$ &16& 17ppt at 20s & 5.98$\times$10$\mathrm{^{-11}}$\\

4.56\cite{nie2024sub} & 6.66$\mathrm{\times10^4}$&65& 0.7ppt at 200s & 3.10$\times$10$\mathrm{^{-11}}$\\

\textbf{4.56 [This work]} & \textbf{2$\mathrm{\times10^3}$} &\textbf{0.035}& \textbf{5ppb at 10s} & \textbf{2.93$\times$10$\mathrm{^{-11}}$}\\

\bottomrule[2pt]
\end{tabularx}
\end{table*}

\newpage

\section{Conclusion}\label{conclusion}

In this work, we showed Lamb-dip resolution spectroscopy and high-sensitivity detection with very low-power operation using a cantilever-based photoacoustic spectroscopy setup targeting a \ce{N2O} trace gas sample. In particular, we demonstrated that simultaneous acoustic amplification and optical cavity enhancement allowed us to operate our PA sensor at low pressure values, down to 60~Pa, enabling for the first time the investigation of non-linear phenomena on trace-gas samples. By showing high signal-to-noise spectra, we proved that such unconventional regimes for acoustic wave generation are still compatible with efficient PA detection. Addressing a strong fundamental \ce{N2O} rovibrational transition, we observed Lamb-dips with an excellent 13\% contrast, in agreement with the theoretical limit, thus showing great potential for high-resolution trace-gas spectroscopy applications. In addition, thanks to the cavity-enhanced configuration, our sensor can perform high-sensitivity trace-gas detection with $\mathrm{\mu}$W-power level. When operated at the optimal sample pressure (i.e., 2$\mathrm{\times 10^3}$~Pa), it achieved a minimum detection limit of \SI{5}{ppb} with an integration time of 10 seconds using an incident optical power as low as \SI{35}{\uW}, corresponding to a NNEA value equal to 2.93$\times$10$\mathrm{^{-11}}$~cm$\mathrm{^{-1}}$WHz$\mathrm{^{-\frac{1}{2}}}$. As far as we know, this is the best NNEA value achieved for a PA-based sensor of $\mathrm{N_2O}$.

Therefore, we demonstrate that, in a single setup, the operation regime can be quickly switched from high-resolution to high-sensitivity using an infrared laser power between 0.035 and 1~mW from a mm-sized QCL. We think that this is a groundbreaking result not only for photoacoustic-based sensors but also for mid-IR precision spectroscopy on molecules. Indeed, if sub-Doppler resolution fills the last gap with respect to the best laser spectroscopy techniques, PA still retains all its intrinsic advantages, namely compactness, wavelength-independent and background-free detection, low-power battery operation, and possible photonic integration. At present, the most advanced spectroscopic techniques can mostly operate in the VIS-NIR spectral region due to the lack of components and detectors of comparable quality in the IR range. However, PA setups do not rely on direct detection of infrared radiation, and this provides easier access to molecules in their fundamental and strongest absorptions. In addition, as a consequence of its zero background detection, Sub-Doppler PA can rely on an ultra-wide dynamic range detection, and further improvements in both the sensitivity and resolution are expected from optimized MEMS transducers and novel mid-IR technologies. These results thus pave the way to breakthrough applications of PA setups for the most advanced molecular sensing applications and for harnessing molecules in the most challenging experiments, like those relying on ultracold samples for fundamental physical and chemical studies, metrology, and quantum technologies.
\newpage

\section*{Supporting information: The acoustic transducer}\label{Supporting}
\begin{figure}[h!]
\centering\includegraphics[width=1\textwidth]{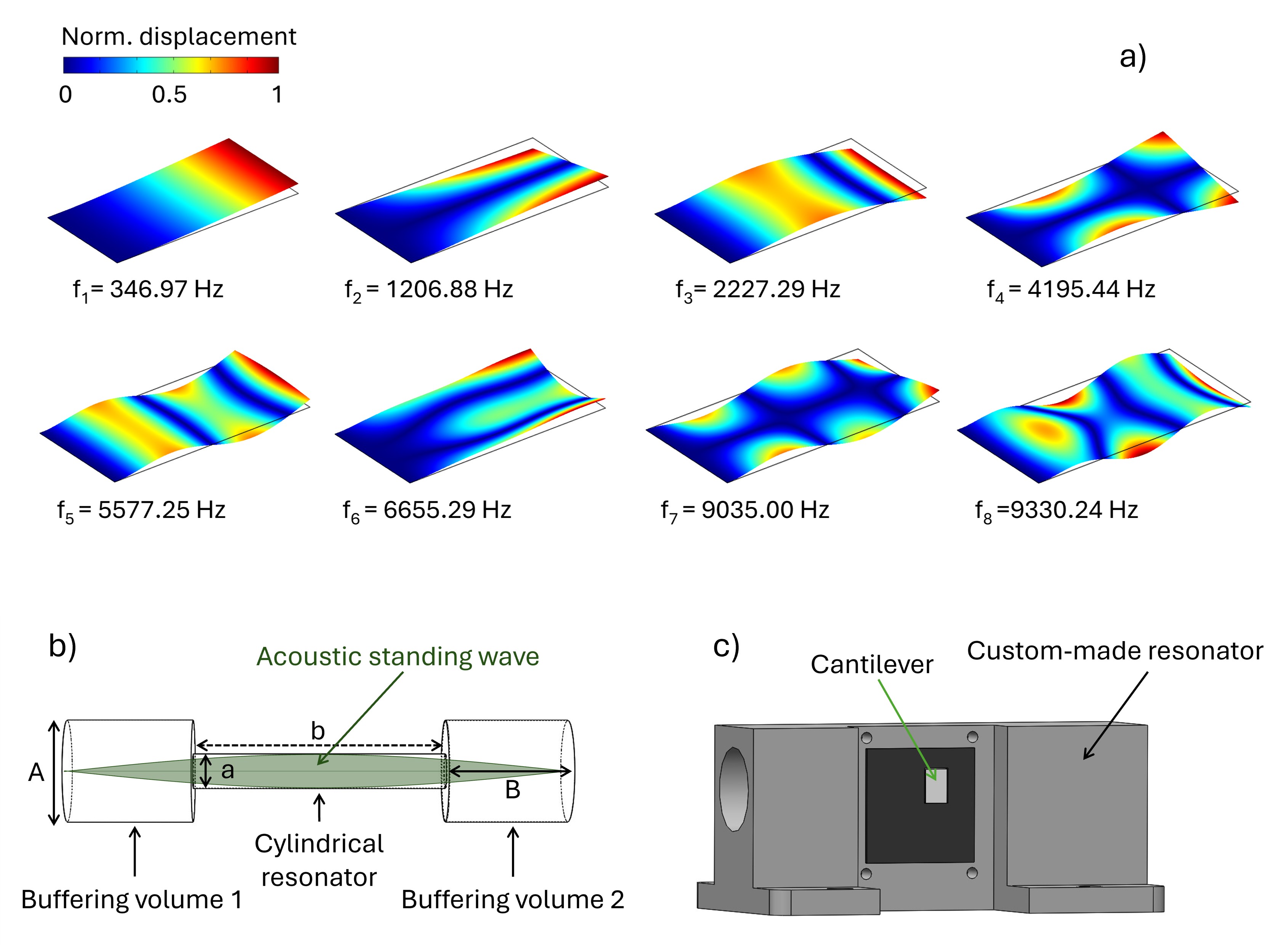}
\caption{\textbf{(a)}: Simulations of the first eight resonance modes of the rectangular cantilever. For each mode, the color map defines the displacement amplitude distribution. The simulations give as results the natural oscillation frequency of the modes. \textbf{(b)}: geometrical features of the developed acoustic resonator. \textbf{(b)}: 3D sketch of the cantilever mounted on the custom-made acoustic resonator}
\label{Cantilver_modes}
\end{figure}

Figure \ref{Cantilver_modes}\textbf{(a)} shows the results of the cantilever's eigenfrequencies analysis, reporting the predicted value of the first eight resonance frequencies and the corresponding modes' displacements. As already discussed in the main text of the manuscript, the cantilever was operated in its fifth resonance. As it is well known from the literature, the confinement of a propagating sound wave within a proper geometrical architecture can induce a significant amplification in its intensity \cite{miklos2001application}. A strategic way to further increase the photoacoustic signal amplitude is to couple the optical resonance (induced by the cavity) with acoustic resonance (achieved employing an acoustic resonator), obtaining a so-called doubly-resonant configuration \cite{wang2022doubly}. According to these assumptions, a custom-made transmission-type acoustic resonator was developed with a geometry theoretically optimized to match the narrow-bandwidth frequency response of the used cantilever.

The device, depicted in Figure \ref{Cantilver_modes}\textbf{(b)}, consists of a cylindrical resonator with length and internal diameter $b=$~~\SI{37}{mm} and $a=$~~\SI{5}{mm} respectively, and two buffering volumes on both sides with a length $B=$~~\SI{18.5}{mm} and diameter $A=$~~\SI{15}{mm}. These dimensions were theoretically selected for optimal acoustic amplification in the 5-kHz frequency range to match the selected cantilever's resonance.
The cantilever is positioned at the resonator's midpoint, corresponding with the anti-node position of the generated standing acoustic wave (depicted in green). To maximize interaction between the cantilever and the impinging acoustic wave in a very reduced volume, the resonator has a small rectangular aperture in correspondence to the cantilever free end, as reported in Figure \ref{Cantilver_modes}\textbf{(c)}.

\medskip

\textbf{Acknowledgements} 
This work was partially supported by EDA under the Cat-B Project Q-LAMPS (EDA contract n.B-PRJ-RT-989), by European Union with the NextGenerationEU Programme I-PHOQS Infrastructure [IR0000016, ID D2B8D520], with the Laserlab-Europe Project [G.A. n.871124], with the MUQUABIS Project [G.A. n.101 070546], by the European Union’s EDF Projects ADEQUADE (ID 101103417) and CASSATA (ID 101121447), by the "National Natural Science Foundation of China (NSFC) (62305279, 52122003)" and by the "Direct Grant for research from The Chinese University of Hong Kong", by the project QOSTRAD (ID 59182/2024) funded by the European Union - Next Generation EU (PNRR-MUR) PE0000023 - NQSTI (National Quantum Science and Technology Institute).

\medskip

\textbf{Data Availability Statement}
The data that support the findings of this study are available from the corresponding author upon reasonable request.

\medskip
\textbf{Conflict of Interest}
The authors declare no conflicts of interest.

\medskip

\bibliography{sample}

@article{black2001introduction,
  title={An introduction to Pound--Drever--Hall laser frequency stabilization},
  author={Black, Eric D},
  journal={American journal of physics},
  volume={69},
  number={1},
  pages={79--87},
  year={2001},
  publisher={American Association of Physics Teachers}
}

@article{patimisco2016analysis,
  title={Analysis of the electro-elastic properties of custom quartz tuning forks for optoacoustic gas sensing},
  author={Patimisco, P and Sampaolo, A and Dong, L and Giglio, M and Scamarcio, Gaetano and Tittel, FK and Spagnolo, V},
  journal={Sensors and Actuators B: Chemical},
  volume={227},
  pages={539--546},
  year={2016},
  publisher={Elsevier}
}

@article{blom1992dependence,
  title={Dependence of the quality factor of micromachined silicon beam resonators on pressure and geometry},
  author={Blom, FR and Bouwstra, S and Elwenspoek, M and Fluitman, JHJ},
  journal={Journal of Vacuum Science \& Technology B: Microelectronics and Nanometer Structures Processing, Measurement, and Phenomena},
  volume={10},
  number={1},
  pages={19--26},
  year={1992},
  publisher={American Vacuum Society}
}

@article{patimisco2014quartz,
  title={Quartz-enhanced photoacoustic spectroscopy: a review},
  author={Patimisco, Pietro and Scamarcio, Gaetano and Tittel, Frank K and Spagnolo, Vincenzo},
  journal={Sensors},
  volume={14},
  number={4},
  pages={6165--6206},
  year={2014},
  publisher={MDPI}
}

@article{haisch2011photoacoustic,
  title={Photoacoustic spectroscopy for analytical measurements},
  author={Haisch, Christoph},
  journal={Measurement Science and Technology},
  volume={23},
  number={1},
  pages={012001},
  year={2011},
  publisher={IOP Publishing}
}

@article{ma2013qepas,
  title={QEPAS based ppb-level detection of CO and N 2 O using a high power CW DFB-QCL},
  author={Ma, Yufei and Lewicki, Rafa{\l} and Razeghi, Manijeh and Tittel, Frank K},
  journal={Optics Express},
  volume={21},
  number={1},
  pages={1008--1019},
  year={2013},
  publisher={Optica Publishing Group}
}

@article{cao2021humidity,
  title={Humidity enhanced N2O photoacoustic sensor with a 4.53 $\mu$m quantum cascade laser and Kalman filter},
  author={Cao, Yuan and Wang, Ruifeng and Peng, Jie and Liu, Kun and Chen, Weidong and Wang, Guishi and Gao, Xiaoming},
  journal={Photoacoustics},
  volume={24},
  pages={100303},
  year={2021},
  publisher={Elsevier}
}

@article{zifarelli2022multi,
  title={Multi-gas quartz-enhanced photoacoustic sensor for environmental monitoring exploiting a Vernier effect-based quantum cascade laser},
  author={Zifarelli, Andrea and De Palo, Raffaele and Patimisco, Pietro and Giglio, Marilena and Sampaolo, Angelo and Blaser, St{\'e}phane and Butet, J{\'e}r{\'e}my and Landry, Olivier and M{\"u}ller, Antoine and Spagnolo, Vincenzo},
  journal={Photoacoustics},
  volume={28},
  pages={100401},
  year={2022},
  publisher={Elsevier}
}

@article{yang2024highly,
  title={Highly sensitive QEPAS sensor for sub-ppb N2O detection using a compact butterfly-packaged quantum cascade laser},
  author={Yang, Min and Wang, Zhen and Sun, Haojia and Hu, Mengyuan and Yeung, Pak To and Nie, Qinxue and Liu, Shanliang and Akikusa, Naota and Ren, Wei},
  journal={Applied Physics B},
  volume={130},
  number={1},
  pages={6},
  year={2024},
  publisher={Springer}
}

@phdthesis{gotti2018comb,
    author = {Gotti, Riccardo},
    title = {Comb-assisted cavity-enhanced molecular spectroscopy at high precision and sensitivity},
    school = {Politecnico di Milano},
    year = {2018}
}

@article{rothman2013hitran2012,
  title={The HITRAN2012 molecular spectroscopic database},
  author={Rothman, Laurence S and Gordon, Iouli E and Babikov, Yury and Barbe, Alain and Benner, D Chris and Bernath, Peter F and Birk, Manfred and Bizzocchi, Luca and Boudon, Vincent and Brown, Linda R and others},
  journal={Journal of Quantitative Spectroscopy and Radiative Transfer},
  volume={130},
  pages={4--50},
  year={2013},
  publisher={Elsevier}
}

@article{pelini2024new,
  title={New silicon-based micro-electro-mechanical systems for photo-acoustic trace-gas detection},
  author={Pelini, Jacopo and Russo, Stefano Dello and Garcia, Inaki Lopez and Canino, Maria Concetta and Roncaglia, Alberto and Pastor, Pablo Cancio and Galli, Iacopo and Ren, Wei and De Natale, Paolo and Wang, Zhen and others},
  journal={Photoacoustics},
  volume={38},
  pages={100619},
  year={2024},
  publisher={Elsevier}
}

@article{hayden2022mid,
  title={Mid-infrared intracavity quartz-enhanced photoacoustic spectroscopy with pptv--Level sensitivity using a T-shaped custom tuning fork},
  author={Hayden, Jakob and Giglio, Marilena and Sampaolo, Angelo and Spagnolo, Vincenzo and Lendl, Bernhard},
  journal={Photoacoustics},
  volume={25},
  pages={100330},
  year={2022},
  publisher={Elsevier}
}

@article{nie2024sub,
  title={Sub-ppt level detection of carbon monoxide and nitrous oxide enabled by mid-infrared doubly resonant photoacoustic spectroscopy},
  author={Nie, Qinxue and Wang, Zhen and Duan, Kun and Hu, Mai and Du, Mengran and Ren, Wei},
  journal={Optics Letters},
  volume={49},
  number={13},
  pages={3648--3651},
  year={2024},
  publisher={Optica Publishing Group}
}

@article{wang2022doubly,
  title={Doubly resonant sub-ppt photoacoustic gas detection with eight decades dynamic range},
  author={Wang, Zhen and Wang, Qiang and Zhang, Hui and Borri, Simone and Galli, Iacopo and Sampaolo, Angelo and Patimisco, Pietro and Spagnolo, Vincenzo Luigi and De Natale, Paolo and Ren, Wei},
  journal={Photoacoustics},
  volume={27},
  pages={100387},
  year={2022},
  publisher={Elsevier}
}

@article{pelini2025cavity,
  title={A cavity-enhanced MEMS-based photoacoustic sensor for ppt-level trace-gas detection},
  author={Pelini, Jacopo and Russo, Stefano Dello and Wang, Zhen and Galli, Iacopo and Pastor, Pablo Cancio and Garcia, Inaki Lopez and Canino, Maria Concetta and Roncaglia, Alberto and Akikusa, Naota and Ren, Wei and others},
  journal={Sensors and Actuators B: Chemical},
  pages={137313},
  year={2025},
  publisher={Elsevier}
}

@misc{patentPAS,
    title={Photoacoustic spectroscopy sensor for trace gas detection and method for trace gas detection},
    author ={Siciliani de Cumis, Mario and Borri, Simone and Canino, Maria Concetta and Cancio Pastor, Pablo and De Natale, Paolo and  Lopez Garcia, Inaki and Roncaglia, Alberto},
    year = {2024},
    month = {feb # "~20"},
    note = {WO2023126455A1}
}

@article{miklos2001application,
  title={Application of acoustic resonators in photoacoustic trace gas analysis and metrology},
  author={Mikl{\'o}s, Andr{\'a}s and Hess, Peter and Boz{\'o}ki, Zolt{\'a}n},
  journal={Review of scientific instruments},
  volume={72},
  number={4},
  pages={1937--1955},
  year={2001},
  publisher={American Institute of Physics}
}

@article{tomberg2019cavity,
  title={Cavity-enhanced cantilever-enhanced photo-acoustic spectroscopy},
  author={Tomberg, Teemu and Hieta, Tuomas and Vainio, Markku and Halonen, Lauri},
  journal={Analyst},
  volume={144},
  number={7},
  pages={2291--2296},
  year={2019},
  publisher={Royal Society of Chemistry}
}

@article{borri2014intracavity,
  title={Intracavity quartz-enhanced photoacoustic sensor},
  author={Borri, S and Patimisco, P and Galli, I and Mazzotti, D and Giusfredi, G and Akikusa, N and Yamanishi, M and Scamarcio, Gaetano and De Natale, P and Spagnolo, V},
  journal={Applied Physics Letters},
  volume={104},
  number={9},
  year={2014},
  publisher={AIP Publishing}
}

@article{russo2024dual,
  title={Dual-tube MEMS-based spectrophone for sub-ppb mid-IR photoacoustic gas detection},
  author={Russo, Stefano Dello and Pelini, Jacopo and Garcia, Inaki Lopez and Canino, Maria Concetta and Roncaglia, Alberto and Pastor, Pablo Cancio and Galli, Iacopo and De Natale, Paolo and Borri, Simone and de Cumis, Mario Sicilani},
  journal={Photoacoustics},
  pages={100644},
  year={2024},
  publisher={Elsevier}
}

@article{di1979sub,
  title={Sub-Doppler optoacoustic spectroscopy},
  author={Di Lieto, Alberto and Minguzzi, Paolo and Tonelli, Mauro},
  journal={Optics Communications},
  volume={31},
  number={1},
  pages={25--27},
  year={1979},
  publisher={Elsevier}
}

@article{rosencwaig1980photoacoustic,
  title={Photoacoustic spectroscopy},
  author={Rosencwaig, Allan},
  journal={Annual review of biophysics and bioengineering},
  volume={9},
  number={1},
  pages={31--54},
  year={1980},
  publisher={Annual Reviews 4139 El Camino Way, PO Box 10139, Palo Alto, CA 94303-0139, USA}
}

@article{kosterev2002quartz,
  title={Quartz-enhanced photoacoustic spectroscopy},
  author={Kosterev, Anatoliy A and Bakhirkin, Yu A and Curl, Robert F and Tittel, Frank K},
  journal={Optics letters},
  volume={27},
  number={21},
  pages={1902--1904},
  year={2002},
  publisher={Optica Publishing Group}
}

@article{wilcken2003optimization,
  title={Optimization of a microphone for photoacoustic spectroscopy},
  author={Wilcken, Klaus and Kauppinen, Jyrki},
  journal={Applied spectroscopy},
  volume={57},
  number={9},
  pages={1087--1092},
  year={2003},
  publisher={Society for Applied Spectroscopy}
}

@article{sun2024multicomponent,
  title={Multicomponent gas sensors based on photoacoustic spectroscopy technology: A review},
  author={Sun, Jialiang and Liu, Bing and Liu, Lixian and Huan, Huiting and Zhang, Le and Zhang, Xueshi and Yang, Junchao and Shao, Xiaopeng},
  journal={Microwave and Optical Technology Letters},
  volume={66},
  number={2},
  pages={e34039},
  year={2024},
  publisher={Wiley Online Library}
}

@article{qiao2024ultra,
  title={Ultra-highly sensitive dual gases detection based on photoacoustic spectroscopy by exploiting a long-wave, high-power, wide-tunable, single-longitudinal-mode solid-state laser},
  author={Qiao, Shunda and He, Ying and Sun, Haiyue and Patimisco, Pietro and Sampaolo, Angelo and Spagnolo, Vincenzo and Ma, Yufei},
  journal={Light: Science \& Applications},
  volume={13},
  number={1},
  pages={100},
  year={2024},
  publisher={Nature Publishing Group UK London}
}

@article{patimisco2015high,
  title={High finesse optical cavity coupled with a quartz-enhanced photoacoustic spectroscopic sensor},
  author={Patimisco, Pietro and Borri, Simone and Galli, Iacopo and Mazzotti, Davide and Giusfredi, Giovanni and Akikusa, Naota and Yamanishi, Masamichi and Scamarcio, Gaetano and De Natale, Paolo and Spagnolo, Vincenzo},
  journal={Analyst},
  volume={140},
  number={3},
  pages={736--743},
  year={2015},
  publisher={Royal Society of Chemistry}
}

@article{nie2023mid,
  title={Mid-infrared swept cavity-enhanced photoacoustic spectroscopy using a quartz tuning fork},
  author={Nie, Qinxue and Wang, Zhen and Borri, Simone and Natale, Paolo De and Ren, Wei},
  journal={Applied Physics Letters},
  volume={123},
  number={5},
  year={2023},
  publisher={AIP Publishing}
}

@article{zheng2024high,
  title={High sensitivity and stability cavity-enhanced photoacoustic spectroscopy with dual-locking scheme},
  author={Zheng, Kaiyuan and Luo, Wenxuan and Duan, Lifu and Zhao, Shuangxiang and Jiang, Shoulin and Bao, Haihong and Ho, Hoi Lut and Zheng, Chuantao and Zhang, Yu and Ye, Weilin and others},
  journal={Sensors and Actuators B: Chemical},
  volume={415},
  pages={135984},
  year={2024},
  publisher={Elsevier}
}

@article{bell1880photophone,
  title={The photophone},
  author={Bell, Alexander Graham},
  journal={Science},
  number={11},
  pages={130--134},
  year={1880},
  publisher={American Association for the Advancement of Science}
}

@book{maddaloni2016laser,
  title={Laser-based measurements for time and frequency domain applications: a handbook},
  author={Maddaloni, Pasquale and Bellini, Marco and De Natale, Paolo},
  year={2016},
  publisher={Taylor \& Francis}
}

@book{letokhov1977nonlinear,
  title={Nonlinear laser spectroscopy},
  volume={4},
  publisher={Springer}
}

@article{wang2025ultrahigh,
  title={Ultrahigh Sensitive LITES Sensor Based on a Trilayer Ultrathin Perfect Absorber Coated T-Head Quartz Tuning Fork},
  author={Wang, Runqiu and Guan, Xueyu and Qiao, Shunda and Jia, Qixiang and He, Ying and Wang, Shaowei and Ma, Yufei},
  journal={Laser \& Photonics Reviews},
  pages={2402107},
  year={2025},
  publisher={Wiley Online Library}
}

@article{meng2025multicomponent,
  title={Multicomponent Gas Measurement Method Based on Miniaturized and Scalable Multiresonator Photoacoustic Spectroscopy},
  author={Meng, Ziqiang and Xiang, Jing and Li, Wei and Xia, Li and Guo, Wenping and Xia, Min and Yang, Kecheng},
  journal={Analytical Chemistry},
  volume={97},
  number={7},
  pages={4158--4165},
  year={2025},
  publisher={ACS Publications}
}

\end{document}